\documentclass[12pt]{article}
\usepackage{epsfig}

\usepackage{amssymb}
\usepackage{amsfonts}

\usepackage{amsthm}
\newtheorem{lem}{Lemma}

\newtheorem{prop}[lem]{Proposition}

\newtheorem{pozn}{Remark}

\def\bp{\begin{proof}}
\def\ep{\end{proof}}

\def\be{\begin{equation}}
\def\ee{\end{equation}}
\def\ba{\begin{array}{c}}
\def\ea{\end{array}}

\def\ben{\[}
\def\een{\]}

\newcommand{\bea}{\begin{eqnarray}}
\newcommand{\eea}{\end{eqnarray}}

\newcommand{\kt}{\rangle}
\newcommand{\br}{\langle}
\begin{document}

\titlepage

\vspace{.35cm}

 \begin{center}{\Large \bf

Solvable non-Hermitian discrete square well with closed-form physical inner product

  }\end{center}

\vspace{10mm}

 \begin{center}

 {\bf Miloslav Znojil}

 \vspace{3mm}
Nuclear Physics Institute ASCR,

250 68 \v{R}e\v{z}, Czech Republic

{e-mail: znojil@ujf.cas.cz}

\vspace{3mm}

%
%\today, chebypol.tex

\end{center}

\vspace{5mm}

%\newpage

\section*{Abstract}

A new Hermitizable quantum model is proposed in which the
bound-state energies are real and given as roots of an elementary
trigonometric expression while the wave function components are
expressed as superpositions of two Chebyshev polynomials. As an
$N-$site lattice version of square well with complex Robin-type
two-parametric boundary conditions the model is unitary with respect
to the Hilbert space metric $\Theta$ which becomes equal to the most
common Dirac's metric $\Theta^{(Dirac)}=I$ in the conventional
textbook Hermitian-Hamiltonian limit. This metric is constructed in
closed form at all $N=2,3,\ldots$.

\subsection*{KEYWORDS}

exactly solvable quantum models;

discrete lattice;

non-Hermitian boundary conditions;

physical inner product;

\newpage

\section{Introduction and summary\label{I} }

Recent studies of quantum bound states in pseudo-Hermitian
representations (cf., e.g., \cite{ali} for a review) revealed the
existence of several paradoxes which are hidden in the conventional
textbook perception of the theory. One of them is that we
almost always select some ``friendly'' Hilbert space ${\cal
H}^{(F)}$ of admissible states (typically,  ${\cal
H}^{(F)}=L^2(\mathbb{R})$ for a point particle) in advance. This
implies that our choice of observables $\Lambda$ (including, in
particular, Hamiltonians $H$) remains {involuntarily restricted}
just to the class of the operators which are self-adjoint {\em in
this particular space}.

Such a model-building strategy has been criticized and rejected by
Bender et al who pointed out that it is far from universal (cf.
review \cite{Carl}). Via a family of Hamiltonians $H$ living in
${\cal H}^{(F)}=L^2(\mathbb{R})$ (with complex $g(x)=(ix)^\delta$
and real $\delta \in (0,2)$ in $H = p^2+g_\delta(x)x^2 \neq
H^\dagger$) these authors demonstrated (and persuaded the other
quantum physicists) that for this type of models the conventional
Hilbert space $L^2(\mathbb{R})$ is clearly {\em unphysical}, false
and ill-chosen.

The resolution of such an $F-$space paradox has been found in
nuclear physics \cite{Dyson,Geyer}. One comes to the conclusion that
whenever $H \neq H^\dagger$ has a real spectrum, the ``first'',
trivial inner product of space ${\cal H}^{(F)}$ (which is called, by
many authors, the ``Dirac's'' inner product) must be declared
unphysical. What should follow is the introduction of an amended,
``sophisticated'' inner-product-metric operator $\Theta^{(S)}$,
i.e., a change of the Hilbert space, ${\cal H}^{(F)} \to {\cal
H}^{(S)}$. Thus, one may keep making calculations in ${\cal
H}^{(F)}$ and upgrade merely the physical expectation values of
observables $\Lambda$ in practice, $\ \br \psi|\Lambda|\psi \kt \to
\br \psi|\Theta \Lambda|\psi \kt$.

The main role of the ``second'' metric $\Theta^{(S)}\neq I$ which
defines the inner product in the physical Hilbert space lies in
making the Hamiltonian {self-adjoint} in ${\cal H}^{(S)}$. The
choice of the Hamiltonian and of the physical Hilbert space {should
proceed in parallel}. The variability of both of the operators $H$
and $\Theta$ is only restricted by the constraint of Hermiticity
imposed upon $H$ in ${\cal H}^{(S)}$, i.e., in the language of space
${\cal H}^{(F)}$, by the condition which has already been known to
Dieudonn\'{e} \cite{Dieudonne},
 \be
 H^\dagger\Theta^{(S)}=\Theta^{(S)}\,H\,.
 \label{dieudoce}
 \ee
Eq.~(\ref{dieudoce}) becomes a guide to a reversal of the
conventional model-building strategy. Newly one starts from a given
$H$ (which is non-Hermitian in unphysical ${\cal H}^{(F)}$) and
inserts this operator and its adjoint in Eq.~(\ref{dieudoce}). Only
then one solves the latter equation for unknown $\Theta^{(S)}$ and
defines the correct physical Hilbert space ${\cal H}^{(S)}$.

% This is also the strategy which will be employed in what follows.

Incidentally, the later recipe leads immediately to another,
$S-$space paradox which has also been mentioned in
Ref.~\cite{Geyer}. Its essence is that a given Hamiltonian may be
assigned {\em many} different metrics $\Theta=\Theta(H)$ via
Eq.~(\ref{dieudoce}). Correctly, the authors of Ref.~\cite{Geyer}
pointed out that such an apparent paradox is in fact purely formal
since a different metric would simply lead to different physics.
With each new phenomenological candidate $\Lambda=\Lambda_j$ for an
observable one has to require its Hermiticity in  ${\cal H}^{(S)}$,
 \be
 \Lambda_j^\dagger\Theta^{(S)}=\Theta^{(S)}\,\Lambda_j
 \,,\ \ \ \ j = 1, 2, \ldots, K
 \,.
 \label{dieudoceje}
 \ee
As long as the metric operator $\Theta^{(S)}$ must remain the same
during the process, one may expect the emergence of an unambiguous
physical Hilbert space ${\cal H}^{(S)}$ at the end.

In our present paper we shall be guided by another, third paradox
which may be noticed to lie in a deep conflict between the required
{\em gain in simplicity} of the Hamiltonian $H \neq H^\dagger$
(highly appreciated, say, in the nuclear-physics applications of
nontrivial metrics \cite{Geyer}) and the parallel {\em loss of
simplicity} of the metrics $\Theta^{(S)} \neq I$. Indeed, although
the latter loss is just the price to be paid for a broadening of the
class of admissible observables (cf. \cite{ali}), the {explicit}
construction of the metric usually appears almost prohibitively
difficult in practice. In this sense it is only necessary to
appreciate that the latter constructions, exact or approximate (cf.,
e.g., Refs.~\cite{171} for a few samples) still have been achieved
for several, often fairly complicated systems in Hilbert spaces of
infinite dimensions (cf., e.g., the remarkable and exceptional
tractability of the ``wrong-sign-oscillator'' in \cite{BG}).

Still, the present-time scarcity of the available metrics in closed
form is certainly one of the weakest points of the theory. For this
reason we decided to search for as elementary an illustration of the
theory as possible, and we found one. First of all, in a way
inspired by the existing extensive literature on linear algebra
(with applicability which is now very well accessible via the
commercially available sophisticated software like MATHEMATICA or
MAPLE) we succeeded because we restricted our choice of models to
the ones represented by the matrices of arbitrary {\em finite}
dimension $N$.

Naturally, we felt inspired also by a lot of work which was already
done, in this direction, in physics (cf., e.g., its small sample in
\cite{chain}, with further references cited therein). In parallel,
we were also guided by the existence of a few ``exceptionally
friendly'' systems in mathematical physics \cite{feasiblesdis} using
linear algebra and matrices of arbitrary (sometimes even doubly
infinite \cite{feasiblesdisbe}) dimension.

Our results will be ordered as follows. First, the selection of the
Hamiltonian $H$ will be discussed in section \ref{II}. For our model
we will find the closed-form solution of Schr\"{o}dinger equation in
section \ref{III}. The existence and localization of the subdomains
${\cal D}$ of physical parameters will be discussed, for which the
spectra of bound state energies remain real and, hence, potentially
observable. In section \ref{IV} we will then concentrate on the key
technical problem of making the system unitary, i.e., on the
constructive search for the missing metrics $\Theta=\Theta(H)$ in
the form of an explicit $N-$parametric solution of
Eq.~(\ref{dieudoce}).

The climax of the story will come with the observation that we shall
be able to make the resulting physical Hilbert space ${\cal
H}^{(S)}$ unique. This goal will be achieved via a less common
though very natural requirement of the smoothness of the parallel
limiting transition of operators $H$ and $\Theta(H)$ to their
respective pre-determined special cases. More explicitly, in a way
preferred by many authors (cf., e.g., \cite{Bila}) we shall demand
and guarantee that whenever the Hamiltonian becomes Hermitian, the
related Hermitizing metric will acquire, in parallel, the
conventional unit-operator Dirac's form of $\Theta^{(Dirac)}=I$.

\section{Preliminaries \label{II}}

\subsection{Differential square wells and their discrete
descendants\label{IIa} }

Ordinary and self-adjoint differential Schr\"{o}dinger equation
 \be
 -\frac{d^2}{dx^2}\psi_n(x)=E_n\psi_n(x)\,,\ \ \ n = 0, 1, \ldots
% x \in (-1,1)\,,\ \ \
 \label{spojham}
 \ee
with the most common square-well boundary conditions
 \be
  \psi_n(\pm 1)=0
 \label{spojhambc}
 \ee
offers one of the most transparent implementations of quantum theory
\cite{Fluegge}. {\it Inter alia}\ it may help in testing
constructive methods and/or in sampling salient mathematical
features of the formalism. In the former, testing context we may,
for example, replace Eq.~(\ref{spojham}) by the sequence of its
discrete, difference-equation descendants living on an equidistant
and left-right symmetric lattice of $N=2M-1$ grid points,
 \ben
 -\psi_n(x_{k-1})+2\,\psi_n(x_{k})-\psi_n(x_{k+1})=
 E_n^{(M)}\psi_n(x_{k})\,,\ \ \
 \ \ \ \
 \ \ \ \
 \een
 \be
 \label{diskrham}
 \ \ \ \
 \ \ \ \
 \ \ \ \
 k = -M+1, -M+2,\ldots,M-1\,.
 \ee
This Schr\"{o}dinger equation may be complemented, say, by the most
common square well boundary conditions
 \be
 \psi_n(x_{\pm M})=0\,.
 \label{diskrhambc}
 \ee
The exact solvability of all of these quantum bound state problems
enables us to compare the $M=\infty$ spectrum of the ordinary
differential Hamiltonian in (\ref{spojham}) + (\ref{spojhambc})
(leading to the ground-state energy $E_0=(\pi/2)^2\approx 2.467$ and
the first excited-state energy $E_1=\pi^2\approx 9.870$ etc) with
the sequences of eigenvalues $E_n^{(M)}$ from Eqs.~(\ref{diskrham})
+ (\ref{diskrhambc}) or rather with their tilded, properly rescaled
values $\tilde{E}_n^{(M)}$ obtained after making the continuous and
discrete systems compatible by fixing the discrete boundary points
$x_{\pm M}=\pm 1$.

A reasonably rapid $M \to \infty$ convergence to the continuous
limit is revealed. With $\tilde{E}_0^{(100)}\approx 2.492$ (and
$\tilde{E}_1^{(100)}\approx 9.968$ etc \cite{disqw}) the error-bar
difference drops below one-percent at $M \approx 100$. Thus, in
methodical setting the choice between differential
Eq.~(\ref{spojham}) + (\ref{spojhambc}) and its discrete analogue
(\ref{diskrham}) + (\ref{diskrhambc}) becomes just a matter of
convenience. At the same time, the study of systems with finite $M$
will open multiple connections with the broad class of the so called
constant-hopping chain models of condensed mater physics
\cite{chain,chainb,chaince}. It is worth adding that the latter,
unexpected correspondence would be lost if we had chosen any
numerically more efficient version of the discretization.

\subsection{Making the differential square well models non-Hermitian\label{IIa}}

The differential-difference and continuous-discrete parallels prove
particularly relevant in the applicability context in which several
interesting results were reported recently. For example, a few
manifestly non-Hermitian upgrades of the square well were proposed
and studied in Refs.~\cite{Langer,sqw,Batal}. In
Ref.~\cite{sdavidem}, attention has been redirected to the specific
model which only lost its Hermiticity (though, incidentally, not the
reality of the bound-state spectrum) due to an {\em ad hoc}
complexification of boundary conditions,
 \be
  \psi(\pm 1)= \frac{\rm i }{\alpha}\,\frac{d}{dx}\psi(\pm 1)\,,
  \ \ \ \ \ \ \ \
 \alpha\, \in \,\mathbb{R}
 %\setminus \pi/2 \mathbb{Z}
  \,.
   \label{spojhambcdav}
 \ee
The exact solvability of the model significantly contributed to the
clarification of multiple open questions \cite{Zelezny}. The
analysis illustrated, in a broader methodical context, what happens
when the Hermiticity of a Hamiltonian is replaced by a weaker though
still phenomenologically tenable assumption of reality (i.e., in
principle, observability) of the bound-state energy levels.

Before proceeding to a core of our present message, viz., to a
deeper analysis of transitions from the differential-operator
Eq.~(\ref{spojham}) to its difference-operator descendant
(\ref{diskrham}) let us add that in Refs.~\cite{Zelezny}, boundary
conditions (\ref{spojhambcdav}) were generalized to read
 \be
  \psi(\pm 1)= \frac{\rm i }{\alpha \mp {\rm i}\beta}
  \,\frac{d}{dx}\psi(\pm 1)\,,
  \ \ \ \ \ \ \ \
 \alpha \,,\, \beta\, \in \,\mathbb{R}
 \,.
   \label{spojhambcdavid}
 \ee
From the point of view of physics, the resulting generalized
non-Hermitian square-well bound-state problem proved much more
interesting since at $\beta=0$ the spectrum was always real. Thus,
only the choice of $\beta > 0$ opened the phenomenologically most
appealing possibility of having phase transitions, i.e., the
spontaneous mergers and subsequent complexifications of colliding
energy pairs. From the point of view of mathematics, unfortunately,
the analysis of the general $\beta\neq 0$ case remained more or less
purely numerical. This fact also contributed to motivation of our
present analysis of the consequences of reduction of continuous
Eq.~(\ref{spojham}) to discrete Eq.~(\ref{diskrham}).

\section{Discrete non-Hermitian square well\label{III}}

\subsection{Discretized boundary conditions
(\ref{spojhambcdavid})\label{IIIb}  }

First of all, let us open the question of replacement of the Robin
boundary conditions (\ref{spojhambcdavid}) by their discrete
equivalent. For this purpose let us  recall the first line of
Eq.~(\ref{diskrham}),
 \ben
 -\psi_n(x_{-M})+2\,\psi_n(x_{-M+1})-\psi_n(x_{-M+2})=
 E_n^{(M)}\psi_n(x_{-M+1})\,
 \een
and the last
line of Eq.~(\ref{diskrham}),
 \ben
 -\psi_n(x_{M-2})+2\,\psi_n(x_{M-1})-\psi_n(x_{M})=
 E_n^{(M)}\psi_n(x_{M-1})\,
 \een
and let us
replace the complex
Robin's differential boundary conditions (\ref{spojhambcdavid})
by their respective first-difference analogues of the respective forms
 \be
  \psi_n(x_{-M})= \frac{\rm i }{\alpha + {\rm i}\beta}\,
  \left (
  \frac{\psi_n(x_{-M+1})-\psi_n(x_{-M})}{h}
  \right )\,,
 \label{prvnibe}
  \ee
 \be
  \psi_n(x_{M})= \frac{\rm i }{\alpha - {\rm i}\beta}\,
  \left (
  \frac{\psi_n(x_{M})-\psi_n(x_{M-1})}{h}
  \right )\,.
  % \label{spojhambcdavid}
 \label{prvnice}
  \ee
Here, a sufficiently small real constant $h>0$ stands for the
difference between the neighboring points of the lattice. The
insertion mediates the elimination of external $\psi_n(x_{-M})$ and
$\psi_n(x_{M})$ and yields
 \be
 \left(
 2-E_n^{(M)}
 -\frac{1}{1 - \beta\,h -i\alpha\,h}
 \right )\,\psi_n(x_{-M+1})-\psi_n(x_{-M+2})=
 0\,,
 \label{prvni}
 \ee
  \be
 -\psi_n(x_{M-2})+
 \left(
 2-E_n^{(M)}
 -\frac{1}{1 - \beta\,h +i\alpha\,h}
 \right )\,\psi_n(x_{M-1})=
 0\,
 \label{druha}
 \ee
i.e., the two final forms of the upgraded,
Robin-boundary-reflecting modification of the respective first and
last line of our present toy-model Schr\"{o}dinger
Eq.~(\ref{diskrham}).

It is worth adding here that besides the present
discretization-approximation derivation and treatment of
Eq.~(\ref{diskrham}) (wherre, incidentally, the value of $z$ is
never purely imaginary), an alternative physical background and
meaning may be also assigned to such a model directly, say, via the
current model-building strategies in condensed matter physics. For
example, in Refs.~\cite{chainb,chaince} (cf. also the related papers
which are cited therein) the authors treat the similar
tridiagonal-matrix models as an $N-$site tight-binding chain. In
such a perspective they decided to work, for simplicity, just with
the two purely imaginary $z$ treated as mutually conjugate
single-site interactions. Even with such a constraint one reveals a
wealth of interesting spectral phenomena. For example, the real
(i.e., stable-system) spectrum may be fragile (cf. \cite{chaince}
and also \cite{fragile} in this respect), while a remarkable
``robust'' exception occurs strictly in the end-point-interaction
special case as studied in the earlier chain-model paper
\cite{chainb}.

\subsection{Three domains of reality of spectra\label{IIIa}  }

The replacement (\ref{spojham}) $\to$ (\ref{diskrham}) of
differential Schr\"{o}dinger equation by its difference-equation
analogue may be perceived as an approximation only in the limit of
very small values of the lattice distance $h = x_{k+1}-x_k$ (or, if
you wish, at the sufficiently large values of the number of grid
points $N = 2M-1$). The same comment also applies to the parallel
replacement of boundary conditions, e.g., (\ref{spojhambcdavid})
$\to$ (\ref{prvnibe}),  (\ref{prvnice}). This indicates that in the
discrete models one could expect the occurrence of a richer spectral
structure, at the larger values of the two variable parameters $\xi
= \alpha h$ and $\zeta=\beta h$ at least.

The most straightforward verification of such a hypothesis may be
performed numerically. Once we abbreviate $1/(1 - \beta\,h
-i\alpha\,h)=z$ and $2-E_n^{(M)} =2y$ we may rewrite our toy-model
Schr\"{o}dinger Eq.~(\ref{diskrham}) + (\ref{prvni}) + (\ref{druha})
in the  compact matrix form
 \be
 \left[ \begin {array}{ccccc}
  2y- z&-1&0
 &\ldots&0
 \\
 \noalign{\medskip}-1&2y&-1&\ddots&\vdots
 \\
 \noalign{\medskip}0&-1&\ddots&\ddots
 &0
 \\
 \noalign{\medskip}\vdots&\ddots&\ddots&2y&-1
 \\
 \noalign{\medskip}0&\ldots&0&-1&2y- z^*
 \end {array} \right]\,
  \left[ \begin {array}{c}
 \phi_1\\
 \noalign{\medskip}\phi_2\\
 \noalign{\medskip}\vdots\\
 \noalign{\medskip}\phi_{N-1}\\
 \noalign{\medskip}\phi_N\\
 \end {array} \right]
 =0\,.
 \label{Ka8}
 \ee
Superscript $^*$ denotes complex conjugation. In order to relax the
artificial constraint $N = 2M-1$ and to admit any $N$ by $N$ matrix
Hamiltonian with $N=2,3,\ldots\,$ we also re-numbered the grid and
the related wave-function values of $\psi_n(x_{-M+k})=\phi_k$.

%\newpage
%********** Figure 0 zde
\begin{figure}[h]                     %instead of \begin{figure}[t]
\begin{center}                         %instead of \begin{center}
\epsfig{file=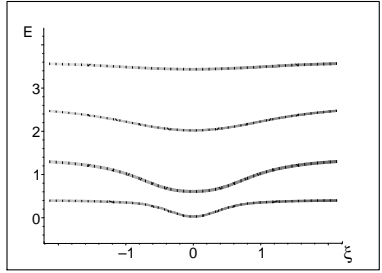,angle=270,width=0.3\textwidth}
\end{center}                         %instead of \end{center}
\vspace{-2mm}\caption{The $\xi-$dependence of the energies $E_n$ for
the $\zeta=0$ Hamiltonian at $N=4$.
 \label{fi01}}
\end{figure}

Even the first, preliminary numerical tests reveal that the
qualitative features of the spectra will vary just smoothly with the
matrix dimension. In particular, Fig.~\ref{fi01} shows that at
$\beta=0$ the unbroken reality of the $N=\infty$ spectrum at all
$\alpha$ \cite{sdavidem} is strictly paralleled even by the first
nontrivial discrete model using $N=4$. This example also shows that
a perceivable effect of the deformation of spectrum of the
conventional Hermitian $\xi \to \infty$ square well ``of the second
kind'' \cite{disqw} (in which $z = 0$) is only obtained at
comparatively small values of $\alpha$ {\em alias} $\xi$.

%\newpage
%********** Figure 0 zde
\begin{figure}[h]                     %instead of \begin{figure}[t]
\begin{center}                         %instead of \begin{center}
\epsfig{file=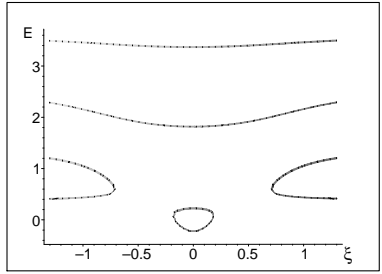,angle=270,width=0.3\textwidth}
\end{center}                         %instead of \end{center}
\vspace{-2mm}\caption{The supercritical $\xi-$dependent
complexification of the first two lowest eigenvalues $E_n$ of the
$\zeta=0.3$ Hamiltonian at $N=4$.
 \label{fi01a}}
\end{figure}

The well known $N=\infty$ observation of the complexification of the
differential-operator spectra beyond certain critical values of
parameters \cite{Tater} is paralleled here by the behavior of the
discretized models equally closely. In particular, beyond certain
$\beta_{critical}>0$, all of the energies only remain real either at
the truly very small values of $\alpha$ {\em alias} $\xi$ (e.g., at
$|\xi|\lessapprox 0.17$ in Fig.~\ref{fi01a} where $N=4$) or in the
two asymptotic, anomalous dynamical regimes of the large $|\xi|$
(e.g., roughly, with $\xi\gtrapprox 0.7$ and $\xi\lessapprox -0.7$
in Fig.~\ref{fi01a}).

%\newpage
%********** Figure 0 zde
\begin{figure}[h]                     %instead of \begin{figure}[t]
\begin{center}                         %instead of \begin{center}
\epsfig{file=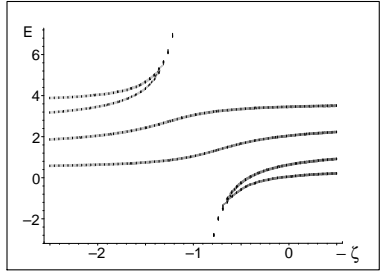,angle=270,width=0.3\textwidth}
\end{center}                         %instead of \end{center}
\vspace{-2mm}\caption{The $(-\zeta)-$dependence of the $N=4$
energies at ${\xi}=0$.
 \label{fi01b}}
\end{figure}
% - mizejici kolecko jde neomezene dolu
%plot3d(tbb4*(1-sep)^2,en4=-5..5,sep=-0.5..2.5,view=0..0.01,
%> orientation=[-180,0.52],axes=boxed,numpoints=4000);

%An explanation of the latter apparent anomaly is easy: the
%deviations from the usual Hermitian square well become very small in
%the latter two large-parameter regimes.

The detailed analysis of the small$-\xi$ spectrum becomes
particularly easy in the limit of $\xi=0$. Its study reveals that
with the restricted growth of $\beta<1/h$ {\em alias} $\zeta<1$ the
allowed interval of small $\xi$ shrinks but still does not vanish.
Surprisingly enough, the doublet of the two lowest real energies
moves down to minus infinity in the limit $\zeta \to 1^-$. At $N=4$
this remarkable feature of the spectrum is documented in
Fig.~\ref{fi01b} which also displays a serendipitous symmetry of the
spectrum with respect to the replacement of $\zeta \to 2-\zeta$.
Naturally, these observations will also find their parallels after
translation into the physical on-site-potential language of the
phenomenological chain models of Refs.~\cite{chainb,chaince}.

%\newpage
%********** Figure 0 zde
\begin{figure}[h]                     %instead of \begin{figure}[t]
\begin{center}                         %instead of \begin{center}
\epsfig{file=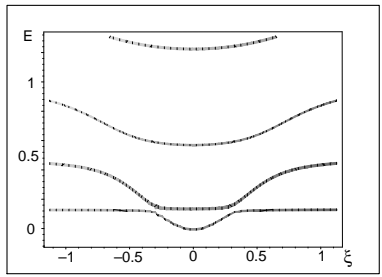,angle=270,width=0.3\textwidth}
\end{center}                         %instead of \end{center}
\vspace{-2mm}\caption{The subcritical $\xi-$dependence
 of the low-lying energies $E_n$ for
the $\zeta=0.05$ Hamiltonian at $N=8$.
 \label{fi02}}
\end{figure}

\begin{figure}[h]                     %instead of \begin{figure}[t]
\begin{center}                         %instead of \begin{center}
\epsfig{file=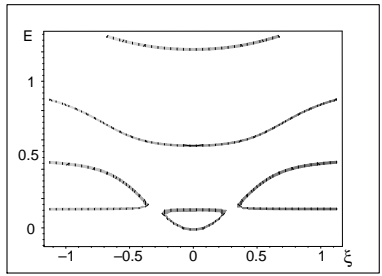,angle=270,width=0.3\textwidth}
\end{center}                         %instead of \end{center}
\vspace{-2mm}\caption{The supercritical $\xi-$dependence
 of the real low-lying energies $E_n$ for
the $\zeta=0.07$ Hamiltonian at $N=8$.
 \label{fi03}}
\end{figure}

With the growth of dimension beyond $N=4$ just inessential,
quantitative variations occur. {\em Before} the breakdown of the
unconditional reality of the spectrum (i.e., at
$\zeta<\zeta_{critical}(N)$) the discrete models exhibit very
similar $\xi-$dependence. During the growth of $|\xi|$ in the
subcritical regime the two lowest energy levels get closer to each
other (see Fig.~\ref{fi02} where we choose $N=8$). Subsequently
(i.e., {\em after} the breakdown, at $\zeta>\zeta_{critical}(N)$),
these two eigenvalues merge and become complex inside two
$N-$dependent intervals of unphysical $\xi$ (cf. Figs.~\ref{fi01a}
or~\ref{fi03} where the complex levels are left invisible).

%\newpage
%********** Figure 0 zde

According to our numerical experiments with emphasis on the regime
with small $\alpha$ {\em alias} $\xi$, the critical value of $\beta$
{\em alias} $\zeta=\zeta_{critical}(N)$ decreases with the
dimension. As an illustration of this observation we may compare,
e.g., $\zeta_{critical}(6)=0.09903$ (which we evaluated, by brute
force, at $N=6$) with Figs.~\ref{fi02} and ~\ref{fi03} which
provide,  at $N=8$, a rough bracketing information about
$\zeta_{critical}(8) \in (0.05,0.07)$.

In the  anomalous regime of the real spectrum at large $|\xi|$, one
can also observe alternative complexifications during a {\em
decrease} of $|\xi|$. In contrast to the previous case, these
complexifications involve the higher excited pairs of levels at the
higher values of $\zeta$. For illustration let us return to the
$N=8$ model for which the small$-\xi$ complexification as shown in
Fig.~\ref{fi03} took place at $\zeta=\zeta_{critical}(8)<0.07 $ and
concerned the lowest levels $E_0$ and $E_1$. We may add that in
contrast to the {\em decrease} of the small$-\xi$ levels $E_0$ and
$E_1$ with the growth of $\zeta$ as sampled in Fig.~\ref{fi01b}, the
asymptotic loss of the reality of the energies during the decrease
of the originally large $|\xi|$ moves upwards. It even involves
higher energy excitations with the growth of $\zeta$. Thus, in the
above $N=8$ example we noticed that near $\zeta=0.5$ the anomalous,
decreasing$-\xi$ complexification is already being transferred from
the first excited pair of $E_1$ and $E_2$ to the second excited pair
of $E_2$ and $E_3$.

\subsection{The exact solvability of the model\label{IIIc}  }

In Eq.~(\ref{Ka8}) let us temporarily omit the first and last line.
The resulting incomplete set
 \be
 \left[ \begin {array}{cccccc}
  -1&2y&-1&0&\ldots&0
 \\
 \noalign{\medskip}0&-1&2y&-1&\ddots
 &\vdots
 \\
 \noalign{\medskip}\vdots&\ddots&\ddots&\ddots&\ddots&0
 \\
 \noalign{\medskip}0&\ldots&0&-1&2y&-1
 \end {array} \right]\,
  \left[ \begin {array}{c}
 \phi_1\\
 \noalign{\medskip}\phi_2\\
 \noalign{\medskip}\phi_3\\
 \noalign{\medskip}\vdots\\
 \noalign{\medskip}\phi_{N}\\
 \end {array} \right]
 =0
 \label{uKa8}
 \ee
of the linear algebraic equations is formally solvable, in closed
form, in terms of Chebyshev polynomials of the first and second kind
\cite{Ryshik},
 \be
 \phi_{n}=A\,T_{n-1}(y)+B\,U_{n-1}(y)\,,\ \ \ \ n = 1, 2, \ldots,
 N\,.
 \label{ansatz}
 \ee
The two free (complex) parameters $A$ and $B$ may be restricted via
a suitable normalization convention, say,
 $
 \phi_{1}=A+B=1$
(naturally, we cannot have a nontrivial solution such that
$\phi_{1}=0$). Thus, we may eliminate $B=1-A$, get formula
 $
 \phi_{2}=(2-A)\,y$
and recall the temporarily omitted first line of Eq.~(\ref{Ka8})
yielding the missing value of the second parameter $A=z/y$ in
Eq.~(\ref{ansatz}), provided only that $y \neq 0$.

The original Schr\"{o}dinger Eq.~(\ref{Ka8}) with $y=0$ must be
considered separately. Taken as recurrences it gives
 \be
 \phi_{2m-1}=(-1)^{m+1}\,,\ \ \
 \phi_{2m}=(-1)^m z\,,\ \ \ m = 1, 2,
 \ldots\,,\ \ \ \ \ y=0\,.
 \label{zeransatz}
 \ee
Incidentally, once these formulae are perceived as the mere
regularized $y\to 0$ limit of Eq.~(\ref{ansatz}), we may conclude
that the wave functions are known in closed form. Now, with the
exceptional $y=0$ the last line of Eq.~(\ref{Ka8}) imposes an
additional constraint upon $z$. At even $N=2K$ it reads $z^*z=1$
while at odd $N = 2K+1$ it has the form $z^*+z=0$.

At odd $N = 2K+1$ the discussion of the latter dichotomy is simpler
because the solutions of the constraint $z^*+z=0= 1-\beta\,h$ may be
omitted as too singular (recall also Fig.~\ref{fi01b} in this
respect). At even $N=2K$ the situation is less complicated. The
assumption of vanishing $y=0$ means that our two free parameters
must satisfy relation $z^*z-1=0=(1-\beta h)^2+\alpha^2h^2-1$ which
determines a circle in the $\alpha-\beta$ plane.

We may summarize that at ``anomalous'' $y=0$ the closed-form
solution of our Schr\"{o}dinger equation is completed. It is worth
adding that at even $N=2K$ and for very small $h$ and not too large
$\alpha$s the circle of parameters degenerates, locally, to a line,
$-2\beta ={\cal O}(h)$. This implies the constancy of $\beta=0$ in
the limit $h \to 0$. Thus, we still stay in a standard dynamical
regime. For example, let us recall that in Fig.~\ref{fi01} where
$K=2$, the second excited level $E_2=2-2y_2$ has strictly vanishing
$y_2=0$ at vanishing $\xi = \alpha\,h=0$.

We may now return to the determination of the other bound-state
energy eigenvalues $y=y_k \neq 0$ for which Schr\"{o}dinger equation
is also reduced to the single algebraic relation represented by the
last line of Eq.~(\ref{Ka8}). After all insertions this
``non-anomalous'' relation acquires the form
 \be
 \frac{z}{y}\,T_{N-2}(y)+\left(
 1-\frac{z}{y}
 \right )\,U_{N-2}(y)=
 (2y-z^*)
 \left [
 \frac{z}{y}\,T_{N-1}(y)+\left(
 1-\frac{z}{y}
 \right )\,U_{N-1}(y)
 \right ]\,
 \ee
admitting simplification
 \be
 -{z}\,U_{N-3}(y)+
 U_{N-2}(y)=
 (2y-z^*)
 \left [
 -z\,U_{N-2}(y)
 +U_{N-1}(y)
 \right ]\,
 \ee
and acquiring the final representation
 \be
 z^*
  z\ U_{N-2}(y)-\left (
 z+z^*
 \right)
 \,U_{N-1}(y)+U_N(y)=0
  \,.
  \label{before}
  \ee
This is secular equation which determines the energy-representing
variables $y_n=1-E_n/2$ as roots of a real polynomial. This means
that these roots are either real or forming complex conjugate pairs.
One can easily check, by direct insertion, that this equation also
leads to the correct answer for the above-discussed (and temporarily
eliminated) ``anomalous'' solutions for which one had $y_n=0$.
Secondly, in the regime of very small parameters  we may set $z
\approx 1$ in Eq.~(\ref{before}) and get
 \be
  U_{N-2}(y_s)-2\,y_s
 \,U_{N-1}(y_s)+U_N(y_s)=2\,(1-y_s)\,U_{N-1}(y_s)=0
  \,.
  \label{rebefore}
  \ee
Thus, the set of roots of this special limiting case is composed of
the separate real item $y_s=1$ and of the $(N-1)-$plet of the well
known and non-degenerate $N-1$ real roots of $U_{N-1}(z)$ which lie
inside interval $(-1,1)$. This enables us to deduce, rigorously,
that a sufficiently small perturbation due to the presence of
non-vanishing parameters $\xi \approx 0$ and $\zeta \approx 0$ will
still keep even the slightly deformed $N-$plet of the roots real
since their possible complexifications may merely proceed via their
pairwise confluences.

In this manner we practically completed the proof of the following
exact-solvability result.

\begin{prop}
\label{lemma1}

For our discrete non-Hermitian square well Schr\"{o}dinger
Eq.~(\ref{Ka8}) the wave functions are given, at any dimension
$N=2,3,\ldots$ and complex $z$, by formula (\ref{ansatz}) or by its
$y \to 0$ limit (\ref{zeransatz}). The reparametrization $y=\cos
\gamma$ of the bound-state energy variables leads to the elementary
trigonometric secular equation
 \be
 z^*
  z\ {\sin (N-1) \gamma}-\left (
 z+z^*
 \right)
 \,{\sin N \gamma}+{\sin (N+1) \gamma}=0
  \,,
  \ \ \ \ 0 \leq \gamma < \pi\,.
  \label{after}
  \ee
Moreover, for the sufficiently small parameters $\xi$ and $\zeta$ in
$z=1/(1-\zeta-i\xi)$, the $N-$plet of the energy roots $y_n=\cos
\gamma_n$ will be real and non-degenerate.

\end{prop}

\bp

Eq.~(\ref{after}) resulted from Eq.~(\ref{before}) after
multiplication by $\sin \gamma$. As long as we may assume that $0
\leq \gamma \leq \pi$, just the endpoints of this interval with
$y=\pm 1$ have to be reconsidered, therefore. Knowing that
$U_n(1)=n+1$ while $U_n(-1)=(-1)^n(n+1)$, the respective special
cases of the correct secular Eq.~(\ref{before}) have the form
 \ben
 (z^*\mp 1)(z\mp 1) N = z^*z-1\,.
 \een
This just restricts the choice of $z$ at a given $N$. More
explicitly, with $\zeta=\beta\,h$ and $\xi=\alpha\,h$ we have
$z=1/(1-\zeta-i\xi)$ so that $zz^*=1/[(1-\zeta)^2+\xi^2]$ and
$z+z^*=2(1-\zeta)/[(1-\zeta)^2+\xi^2]$.

In the first step we have to re-analyze our correct secular equation
at $y=1$,
 \ben
  \frac{N-1}{(1-\zeta)^2+\xi^2}-\frac{2N(1-\zeta)}{(1-\zeta)^2+\xi^2}
  +N+1=0\,.
 \een
This leads to the elimination of
 \ben
 \xi^2=-\zeta^2+\frac{2\zeta}{N+1}
 \een
so that our $y=1$ solution [i.e., the obviously existing $\gamma=0$
solution of Eq.~(\ref{after})] will be acceptable whenever
 \ben
 0 \leq \zeta \leq \frac{2}{N+1}\,.
 \een
This is compatible with our parameter-smallness assumptions.

In the second step we set $y=-1$ and obtain the constraint
 \ben
  \frac{N-1}{(1-\zeta)^2+\xi^2}+\frac{2N(1-\zeta)}{(1-\zeta)^2+\xi^2}
  +N+1=0\,.
 \een
For positive $\xi=\rho \sin \delta$ and $\zeta =\rho \cos \delta$
with $0\leq \delta \leq \pi/2$ this enables us to eliminate
 \ben
 \sin \delta = \frac{4N+(N+1)\rho^2}{(4N+2)\rho}
 \een
and treat $\rho$ as a free parameter which keeps the angle $\delta$
real if and only if
 \ben
 2-\frac{2}{N+1} \leq \rho \leq 2\,.
 \een
We see that either $\xi$ or $\zeta$ cannot be small in such a case.
Thus, by inequality $\gamma<\pi$ the artificially added $(N+1)-$st
root $\gamma=\pi$ of Eq.~(\ref{after}) was crossed out again as
redundant, in the small-parametric dynamical regime at least.

\ep

\section{The metrics \label{IV}}

At any matrix dimension $N<\infty$ and for the physical parameters
guaranteeing the reality of the spectrum of $H$ the determination of
{\em any} acceptable (i.e., Hermitian and positive-definite) matrix
solution $\Theta$ of Dieudonn\'{e}'s Eq.~(\ref{dieudoce}) is rarely
feasible and always ambiguous. Even in the feasible cases (cf.,
e.g., \cite{feasibles}) the knowledge of an output of virtually any
preliminary symbolic-manipulation calculation hardly helps and often
provides just long and illegible formulae. This makes the
constructions of the metrics rather challenging even at the very
small dimensions $N$.

\subsection{All the metrics at $N=3$\label{IVa}}

For the sake of simplicity let us just consider the simplest dynamical
$\zeta=0$ scenario and Hamiltonian
 \be
 H=\left[ \begin {array}{ccc} - \left( 1-i{\it \xi} \right) ^{-1}&-1&0
\\\noalign{\medskip}-1&0&-1\\\noalign{\medskip}0&-1&- \left( 1+i{\it
\xi} \right) ^{-1}\end {array} \right]\,.
 \ee
All of the possible related metrics may be sought in the form of a
six-parametric complex ansatz
 \be
 \Theta=
 \left[ \begin {array}{ccc} r&u+i{\it u_2}&
 {\it z_1}+i{\it z_2}\\\noalign{\medskip}u-i{\it u_2}
 &s&u+i{\it u_2}\\\noalign{\medskip}{\it
 z_1}-i{\it z_2}&u-i{\it u_2}&r\end {array} \right]\,.
 \label{ounz}
 \ee
Its insertion converts Dieudonn\'{e}'s Eq.~(\ref{dieudoce})
into a set of nine (not necessarily independent) relations
with a complete three-parametric solution
 \ben
 {\it u_2}=-{\frac {r \, \xi  }{1+{  \xi  }^{2}}}\,,
 \ \ \ \ \ \ \ \
% \ee
%
%
% \be
%z2= \xi*(s-z1-2*r-z1*\xi^2-r*\xi^2+s*\xi^2)/(1+\xi^2)
%{\it z2}={\frac {\xi\, \left( s-{\it z1}-2\,r-{\it z1}\,{\xi}^{2}-r{
%\xi}^{2}+s{\xi}^{2} \right) }{1+{\xi}^{2}}}
{\it z_2}=-{\frac { \left( u{\xi}^{2}+r+u \right) \xi}{ \left( 1+{\xi}^{2}
 \right) ^{2}}}
 \een

 \ben
%z1=(s-r+u+2*s*\xi^2-3*r*\xi^2-\xi^4*r+\xi^4*s+u*\xi^2)/(1+2*\xi^2+\xi^4);
{\it z_1}={\frac {s-r+u+2\,s{\xi}^{2}-3\,r{\xi}^{2}-{\xi}^{4}r+{\xi}^{4
}s+u{\xi}^{2}}{\left( 1+{\xi}^{2}
 \right) ^{2}}}\,.
 \een
Obviously, even though we choose just the first nontrivial dimension
$N=3$, an exhaustive discussion of all of the resulting optional
kinematical scenarios looks prohibitively complicated. In fact, it
need not be necessarily so: a few constructive samples of the
analysis of what happens to the complete sets of operators of
observables under nontrivial metrics may be found, e.g., in
\cite{sample}.

Whenever $N \geq 3$, a practically necessary guide to
simplifications of our present constructions of metrics has been
found in the idea that whenever the Hamiltonian $H$ becomes
self-adjoint in ${\cal H}^{(F)}$ at some special parameters (i.e.,
in the limit of $\xi \to 0$ in our case), all of the metrics
$\Theta$ which remain, in the same limit, different from the
``natural'' Dirac's trivial $\Theta^{(F)}=I$ might be discarded as
anomalous and pathological. On this basis we set $s=r=1$ and got the
special, simplified one-parametric family of the $N=3$ metrics
possessing a nicer and more compact form
 \be
 \Theta(u)=\left[ \begin {array}{ccc} 1&u
 -{\frac {i{\it {\xi}}}{1+{{\it {\xi}}}^{2}}}&
{\frac {u-{{\it {\xi}}}^{2}+u{{\it {\xi}}}^{2}}{1+2\,{{\it
{\xi}}}^{2}+{{\it {\xi} }}^{4}}}-{\frac {i \left( u{{\it
{\xi}}}^{2}+1+u \right) {\it {\xi}}}{
 \left( 1+{{\it {\xi}}}^{2} \right) ^{2}}}\\\noalign{\medskip}u+{\frac {i
{\it {\xi}}}{1+{{\it {\xi}}}^{2}}}&1&u-{\frac {i{\it {\xi}}}{1+{{\it
{\xi}}}^{2}}}
\\\noalign{\medskip}{\frac {u-{{\it {\xi}}}^{2}+u{{\it {\xi}}}^{2}}{1+2\,{{
\it {\xi}}}^{2}+{{\it {\xi}}}^{4}}}+{\frac {i \left( u{{\it
{\xi}}}^{2}+1+u
 \right) {\it {\xi}}}{ \left( 1+{{\it {\xi}}}^{2}
  \right) ^{2}}}&u+{\frac {i
{\it {\xi}}}{1+{{\it {\xi}}}^{2}}}&1\end {array} \right]\,.
\label{metr3}
 \ee
This matrix is amenable to numerical testing. One reveals that for
any $u\neq 0$, the ${\xi}\to 0$ limit of such a matrix will differ
from the isotropic Dirac's $\Theta^{(F)}=I$. For this reason we
finally set $u=0$. The resulting matrix $\Theta(0)$ proved tractable
as a metric because it remains positive definite in the whole
physical (i.e., real-energies-guaranteeing) range of parameter $\xi
\in (-\infty,\infty)$.

%\newpage
%********** Figure 0 zde
\begin{figure}[h]                     %instead of \begin{figure}[t]
\begin{center}                         %instead of \begin{center}
\epsfig{file=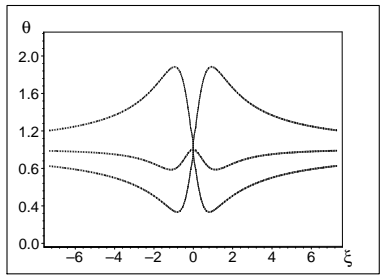,angle=270,width=0.3\textwidth}
\end{center}                         %instead of \end{center}
\vspace{-2mm}\caption{The ${\xi}-$dependence of eigenvalues of the
$N=3$ metric (\ref{ounz}) with $s=r=1$ and $u=0$.
 \label{fi06}}
\end{figure}

Fig.~\ref{fi06} displays the ${\xi}-$dependence of the triplet of
the metric's eigenvalues. One should add that up to a constant
factor, the smoothness of transition $\Theta(0)\to \Theta^{(F)}=I$
happens to be guaranteed not only in the ``natural''
Hermitian-Hamiltonian limit of small $\xi \to 0$ but also, beyond
the scope of Fig.~\ref{fi06}, in the other two Hermitian-Hamiltonian
limits $\xi \to \pm \infty$. In this sense metric $\Theta(0)$ may be
considered optimal.

\subsection{Special metric at $N=4$\label{IVb}}

Along the same lines as in preceding subsection one may insert the
adjusted general $N=4$ ansatz for $\Theta$ and insert it again in
Eq.~(\ref{dieudoce}). The computer-assisted symbolic-manipulation
solution of this set of $N^2$ linear algebraic equations is routine.
It yields the formulae (too long for a display here) which contain
now as many as four independent free real parameters.

\begin{figure}[h]                     %instead of \begin{figure}[t]
\begin{center}                         %instead of \begin{center}
\epsfig{file=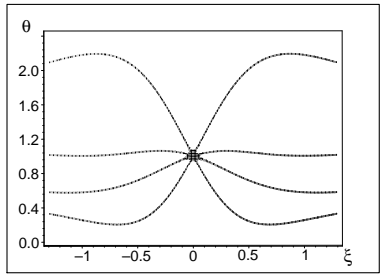,angle=270,width=0.3\textwidth}
\end{center}                         %instead of \end{center}
\vspace{-2mm}\caption{The eigenvalues of metric (\ref{simpli}) with
$u_1=0$.
 \label{fi07}}
\end{figure}

It proved rather fortunate that the extrapolated, parameter-free
simplified ansatz
 \be
 \Theta=
\left[ \begin {array}{cccc} 1&{\it u_1}+i{\it u_2}&{\it z_1}+i{\it
z_2}&{\it r_1}+i{\it r_2}\\\noalign{\medskip}{\it u_1}-i{\it
u_2}&1&{\it u_1}+i{ \it u_2}&{\it z_1}+i{\it
z_2}\\\noalign{\medskip}{\it z_1}-i{\it z_2}&{\it u_1}-i{\it
u_2}&1&{\it u_1}+i{\it u_2}\\\noalign{\medskip}{\it r_1}-i{\it
r_2}&{\it z_1}-i{\it z_2}&{\it u_1}-i{\it u_2}&1\end {array} \right]
\label{simpli}
 \ee
with units on its main diagonal appeared optimal again. Not too
surprisingly, the limiting ${\xi}\to 0$ case would remain
anisotropic for $u_1\neq 0$. Nevertheless, the choice of $u_1=0$
re-established the isotropy. Moreover, the related simplification
left the result of the symbolic-manipulation solution of
Eq.~(\ref{dieudoce}) printable, yielding
 \ben
 u_2=-{\frac {{\it {\xi}}}{1+{{\it {\xi}}}^{2}}}\,,
 \ \ \ \ \ \
 %
% \een
%
% \ben
 z_1=-{\frac {{{\it {\xi}}}^{2}}{\left( 1+{{\it {\xi}}}^{2}
\right) ^{2}}}\,,
 \ \ \ \ \ \
% \een
% \ben
 z_2=-{\frac {{\it {\xi}}}{ \left( 1+{{\it {\xi}}}^{2}
\right) ^{2}}}\,,
 \ \ \
 \een
 \be
 r_1=-2\, \frac {{{\it {\xi}}}^{2}}{{\left( 1+{{\it {\xi}}}^{2}
\right) ^{3}}}\,,
 \ \ \ \ \ \
 %
% \een
% \ben
 r_2=-{\frac {
\left( -{{\it {\xi}}}^{2}+1 \right) {\it {\xi}}}{   \left( 1+{{\it
{\xi}}}^{2}
 \right)^3 }}\,.
 \label{rema}
 \ee
Last but not least, it was important to find that also all the four
eigenvalues of the resulting unique metric remained positive and
finite in the whole physical range of the dynamical parameter
${\xi}\in (-\infty,\infty)$ and that they all proved equal in all of
the three alternative Hermitian limits of $\xi \to 0$ and  $\xi \to
\pm \infty$ (cf. Fig.~\ref{fi07}).

%\newpage
%********** Figure 0 zde

\subsection{A reparametrization of Hamiltonians}

%\subsection{$N=3$}

One of the remarkable features of matrix elements (\ref{rema}) is
that they vanish in all of the three Hermiticity limits, i.e., in
the extremes of both small and large parameters $\xi$. Naturally,
one pays for the universality by the complicated algebraic form of
these formulae.

There exist several reasons why one should not pay too much
attention to the two rather artificial large$-\xi$ physical domains.
Naturally, they do not have any direct $N \to \infty$ limiting
connection with the continuous Schr\"{o}dinger Eq.~(\ref{spojham})
with the complex Robin boundary conditions (\ref{spojhambcdav}) or
(\ref{spojhambcdavid}). In this sense we shall assume, from now on,
that both of the parameters $\xi$ and $\zeta$ in the Hamiltonian
$H(\xi,\zeta)$ of Eq.~(\ref{Ka8}) remain small and that they may be
replaced by another real and small pair, say, $(\omega,\varrho)$,
with the maximal attention paid to the resulting simplification of
the matrices in question.

As long as the realization of such a trick involves just the two
extreme elements in our Hamiltonians, it is sufficient to illustrate
it at $N=3$. Thus, we decided to require that
 \be
 H=\left[ \begin {array}{ccc} -z(\zeta,\xi)&-1&0\\
 \noalign{\medskip}-1&0&-1\\
 \noalign{\medskip}0&-1&-z(\zeta,-\xi)\end {array} \right]
 =
\left[ \begin {array}{ccc} -1-\varrho-i{\it
 {\omega}}&-1&0\\
 \noalign{\medskip}-1&0&-1\\
 \noalign{\medskip}0&-1&-1-\varrho+i{\it {\omega}}\end {array}
 \right]\,
 \label{rambousek}
 \ee
i.e.,
 \be
 z(\zeta,\xi)=\frac{1}{1-\zeta-i\xi}= 1-\varrho+i\omega\,.
 \label{reim}
 \ee
An easy algebra leads to the formulae
 \be
 \omega=\omega(\zeta,\xi)=\frac{\xi}{(1-\zeta)^2+\xi^2}\,,
 \ee
 \be
 \varrho=\varrho(\zeta,\xi)=
 \frac{\zeta-\zeta^2-\xi^2}{(1-\zeta)^2+\xi^2}\,.
 \label{rovproro}
 \ee
Although such a reparametrization $(\xi,\zeta) \to (\omega,\varrho)$
of the whole real plane is rather complicated globally, we simply
have the trivial mapping $\omega \sim \xi$ and $\varrho \sim \zeta$
in one of the most interesting physical domains near the origin.
Secondly, in the condensed-matter context of
Refs.~\cite{chainb,chaince} our reparametrization (\ref{reim}) of
the interaction may be also perceived as establishing a more direct
connection between our mathematical results and the physics of the
{\em stable} ${\cal PT}-$symmetric condensed-matter $N-$site
lattices containing a pair of weak impurities localized at the two
ends.

%\newpage
%********** Figure 0 zde
\begin{figure}[h]                     %instead of \begin{figure}[t]
\begin{center}                         %instead of \begin{center}
\epsfig{file=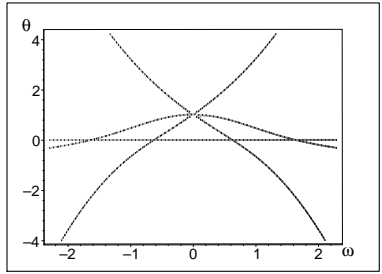,angle=270,width=0.3\textwidth}
\end{center}                         %instead of \end{center}
\vspace{-2mm}\caption{Eigenvalues of metric (\ref{rambou}) at
$\varrho=0$. The loss of positivity with the growth of $|\omega|$.
 \label{8fi0}}
\end{figure}

Besides the natural physical and perturbation-theory motivated
interest in the parametric domain near the origin, the decisive
advantage of our reparametrization lies in the {\em enormous}
simplification of the related {\em exact} matrices of the metrics.
For example, at $\varrho=0$ the right-hand-side Hamiltonian matrix
of Eq.~(\ref{rambousek}) is now being assigned the exact $u=0$
Hermitizing metric
 \be
 \Theta^{(3)}=
 \left[ \begin {array}{ccc} 1{}&-i{\it {\omega}}
 &-{{\it {\omega}}}^{2}-i{
 \it {\omega}}\\
 \noalign{\medskip}i{\it {\omega}}&1{}&-i{\it {\omega}}
 \\
 \noalign{\medskip}-{{\it {\omega}}}^{2}+i{\it {\omega}}&i{\it
 {\omega}}&1{}
 \end {array} \right]\,.
 \label{rambou}
 \ee
A routine check confirms that at the small values of $\omega$ this
matrix is positive definite as it should be. According to
Fig.~\ref{8fi0} this matrix ceases to be positive definite at
certain larger $\omega_{critical}$ but this just reflects the fact
that the energy spectrum of the Hamiltonian itself ceases to be real
at these points which lie comparatively far from the origin. One
should add that this does not contradict Figs.~\ref{fi01}
and/or~\ref{fi06}, either, because due to Eq.~(\ref{rovproro}) the
line of $\varrho=0$ is represented by a comparatively large circle
in the $\zeta-\xi$ plane.

The even more persuasive advantages of our reparametrization emerge
during a transition to the larger dimensions $N$. Thus, the next,
$N=4$ and $\varrho=0$ Hamiltonian
 $$
 H=\left[ \begin {array}{cccc} -1-i{\it \omega}
 &-1&0&0\\
 \noalign{\medskip}-1&0&-1&0\\
 \noalign{\medskip}0&-1&0&-1\\
 \noalign{\medskip}0&0&-1&-1+i{
\it \omega}\end {array} \right]
 $$
is being assigned the $u=0$ metric
 \be
 \Theta^{(4)}=
 \left[ \begin {array}{cccc} 1{}&-i{\it \omega}&-{{\it \omega}}^{2}-i{
\it \omega}&-2\,{{\it \omega}}^{2}+i \left( {{\it \omega}}^{3}-{\it
\omega} \right)
\\
\noalign{\medskip}i{\it \omega}&1{}&-i{\it \omega}&-{{\it
\omega}}^{2}-i{
\it \omega}\\
\noalign{\medskip}-{{\it \omega}}^{2}+i{\it \omega}&i{\it
\omega}&1{}&-i{\it \omega}\\
\noalign{\medskip}-2\,{{\it \omega}}^{2}-i \left( {{\it \omega
}}^{3}-{\it \omega} \right) &-{{\it \omega}}^{2}+i{\it \omega}&i{\it
\omega}&1-{\it th }\end {array} \right]
 \label{ram4}
 \ee
with the spectrum which is fully analogous to the one of
Fig.~\ref{8fi0} above.

\subsection{Metrics at all matrix dimensions $N$}

On the background of preceding preparatory considerations there
emerges a pattern which may be made explicit and formulated as our
second main result.

\begin{prop}

\label{lemma2}

For the special and reparametrized one-parametric $\varrho=0$ subset
 \be
 H^{(N)}(\omega)=\left[ \begin {array}{ccccc}
 -1-i{\it \omega} &-1&0&\ldots&0\\
 \noalign{\medskip}-1&0&-1&\ddots&\vdots\\
 \noalign{\medskip}0&\ddots&\ddots&\ddots&0\\
 \noalign{\medskip}\vdots&\ddots&-1&0&-1\\
 \noalign{\medskip}0&\ldots&0&-1&-1+i{\it \omega}
 \end {array} \right]
 \label{uMETen}
 \ee
of our $N$ by $N$ matrix Hamiltonians of Eq.~(\ref{Ka8}) with $N
\geq 2$ the closed form of Hermitizing metric $\Theta^{(N)}(\omega)$
with the Dirac-limit property $\Theta^{(N)}(0)=I$ is given by
formulae $\Theta_{nn}^{(N)}=1$, $n=1,2,\ldots,N$ and
 \be
  \Theta_{nn+k}^{(N)}=
  \left (\Theta_{n+k,n}^{(N)}\right )^*
  =-i\omega\,(1-i\omega)^{k-1}\,,
  \ \ \ \ n =  1, 2, \ldots, N-k\,,\ \ \ \
 k = 1, 2, \ldots, N-1\,.
 \label{METn}
 \ee
\end{prop}

\bp

The choice of $\Theta_{nn}^{(N)}=1$ with $n=1,2,\ldots,N$ fixes some
of the available free parameters so that we just have to prove the
validity of formulae (\ref{METn}). This result is trivial at $N=2$
and it has been shown to hold at $N=3$ and $N=4$ in the preceding
text (cf. Eqs.~(\ref{rambou}) and (\ref{ram4}), respectively). It
remains for us to show that the metric $\Theta^{(N)}(\omega)$ and
matrix $H$ of Eq.~(\ref{uMETen}) are mutually compatible with
respect to the Dieudonn\'{e}'s constraint (\ref{dieudoce}) at any
$N$.

For our present purposes the latter constraint may be perceived as
the $N$ by $N$ matrix set of linear algebraic equations ${\cal
M}_{ij}^{(N)}=0$. Thus, starting from a generic ansatz
 \ben
  \Theta_{nn}^{(N)}=
  \left (\Theta_{nn}^{(N)}\right )^*
  =p^{(1)}_1+ i\,p^{(1)}_2\,, \ \ \ \
  \Theta_{nn+1}^{(N)}=
  \left (\Theta_{n+1,n}^{(N)}\right )^*
  =p^{(2)}_1+ i\,p^{(2)}_2\,,\ \ \ldots
  \een
  \ben
  \ \ \ \ \ldots\,
 \Theta_{n,N}^{(N)}=
  \left (\Theta_{N,n}^{(N)}\right )^*
  =p^{(N)}_1+ i\,p^{(N)}_2\,
 \label{seMETn}
 \een
we immediately see that the knowledge of the ``old'' matrix elements
$p^{(N-j)}_1+ i\,p^{(N-j)}_2$ with $j=1,2,\ldots,N-1$ leaves just
the following four nontrivial complex relations in
Eq.~(\ref{dieudoce}),
 \ben
 {\cal M}_{ij}^{(N)}=0\,,\ \ \ \
 (i,j) \in \{
 (1,N-1), (2,N), (N-1,1), (N,2) \}\,.
 \een
They are, incidentally, linearly dependent. Thus, taking any one of
these relations, we have to split it into its real and imaginary
parts. This yields the pair of the real recurrence relations
 \be
 p_1^{(N)}=p_1^{(N-1)}+\omega\,p_2^{(N-1)}\,,\ \ \ \ \
 p_2^{(N)}=p_2^{(N-1)}-\omega\,p_1^{(N-1)}\,.
 \label{tayre}
 \ee
A patient mathematical induction may then start from the known
initial values of
 \ben
 p_1^{(1)}=1\,,\ \ \ \ \
 p_2^{(1)}=0\,
 \een
or, if you wish, from one of the above-mentioned
explicit-calculation results with
 \ben
 p_1^{(2)}=0\,,\ \ \ \ \
 p_2^{(2)}=-\omega\,,\ \ \ \
 p_1^{(3)}=-\omega^2\,,\ \ \ \ \
 p_2^{(3)}=-\omega\,
 \een
etc. The elementary combinatorics finally proves the step $N-1 \to
N$.

\ep

\begin{pozn}
From the formal point of view, the restriction of our attention to
the metrics with all of the free parameters fixed by the condition
$\Theta^{(N)}(0)=I$ may still have many well founded alternatives.
We would like to add that in the case of need of any more flexible
family of metrics, the extreme simplicity of our Hamiltonians may be
expected to admit still some closed-form alternative constructions
in which $\Theta^{(N)}(0)\neq I$.
\end{pozn}
Their first sample is
the following

\begin{prop}

\label{lemma3}

The same $N$ by $N$ matrix Hamiltonians (\ref{uMETen}) as above may
be assigned the closed-form Hermitizing metric
$\Theta^{(N)}(\omega)=\Theta^{(N)}_{(0)}(\omega)+u\,\Theta^{(N)}_{(1)}(\omega)$
where $\Theta^{(N)}_{(0)}(\omega)$ denotes the $u=0$ metric of
Proposition \ref{lemma2}. Closed formulae define  $\left
(\Theta_{(1)} \right )_{nn}^{(N)}=0$, $n=1,2,\ldots,N$ and
 \be
  \left (\Theta_{(1)}  \right )_{nn+k}^{(N)}=
  \left [\left (\Theta_{(1)}  \right )_{n+k,n}^{(N)}\right ]^*
  =(1-i\omega)^{k-1}\,,
  \ \ \ \ n =  1, 2, \ldots, N-k\,,\ \ \ \
 k = 1, 2, \ldots, N-1\,.
 \label{METan}
 \ee
At each $N$ the value of parameter $u$ must remain sufficiently
small to keep the metric positive definite.

\end{prop}

\bp

Naturally, once we use the amended guess of the form of matrix
elements of metric, its direct insertion is easily shown to convert
also the above-mentioned non-trivial components of the set of
Dieudonnian recurrences (\ref{tayre}) into identities.

 \ep

In an alternative to the above proof, we could still use an entirely
general ansatz and proceed by mathematical induction as above. Even
in such a more traditional approach the verification of
compatibility between metric and Hamiltonian may be facilitated by
the use of computer-assisted symbolic manipulations.

%
%%\vspace{15mm}
%
%\section*{Acknowledgement}
%
%Work supported by the
%%GA\v{C}R grant Nr. P203/11/1433, by the
%%M\v{S}MT ``Doppler Institute" project Nr. LC06002 and by the
%Institutional Research Plan AV0Z10480505.
%
%
%


\newpage




\begin{thebibliography}{00}

\bibitem{ali}
A. Mostafazadeh,
%Pseudo-Hermitian Quantum Mechanics,
%arXiv:0810.5643,
%%.
%% arXiv:0810.5643 [pdf, ps, other]
%%Title: Pseudo-Hermitian Representation of Quantum Mechanics Authors:
%%Ali Mostafazadeh Comments: 76 pages, 2 figures, 243 references,
%%revised version
%to appear in Int. J. Geom. Meth. Mod. Phys.
%% Subjects:
%%Quantum Physics (quant-ph); High Energy Physics - Theory (hep-th);
%%Mathematical Physics (math-ph)
%Mostafazadeh, A.: Pseudo-Hermitian Representation of Quantum Mechanics.
Int. J. Geom. Meth. Mod. Phys. 7 (2010) 1191.
%1306

\bibitem{Carl}
C. M. Bender, Rep. Prog. Phys. 70 (2007) 947.


\bibitem{Dyson}
F. J. Dyson,
%General Theory of Spin-Wave Interactions,
Phys. Rev. 102 (1956) 1217.
%-1230,.

\bibitem{Geyer}
F. G. Scholtz, H. B. Geyer and F. J. W. Hahne, Ann. Phys. (NY) 213
 (1992) 74.
%% Quasi-Hermitian Operators in Quantum Mechanics
%% and the Variational Principle,


\bibitem{Dieudonne}
J. Dieudonne,
%Quasi-Hermitian operators,
Proc. Int. Symp. Lin. Spaces, Pergamon, Oxford,
1961, pp. 115-122.

\bibitem{171}
M. S. Swanson, J. Math. Phys. 45 (2004) 585;
%
%
% 134. Transition elements for a non-Hermitian quadratic Hamiltonian
%By: Swanson, MS JOURNAL OF MATHEMATICAL PHYSICS  Volume: 45   Issue:
%2   Pages: 585-601   Published: FEB 2004
% Full Text from Publisher Full Text View AbstractClose
%
% Abstract
%The non-Hermitian quadratic Hamiltonian
%H=omegaa(dagger)a+alphaa(2)+betaa(dagger2) is analyzed, where
%a(dagger) and a are harmonic oscillator creation and annihilation
%operators and omega, alpha, and beta are real constants. For the
%case that omega(2)-4alphabetagreater than or equal to0, it is shown
%using operator techniques that the Hamiltonian possesses real and
%positive eigenvalues. A generalized Bogoliubov transformation allows
%the energy eigenstates to be constructed from the algebra and states
%of the harmonic oscillator. The eigenstates are shown to possess an
%imaginary norm for a large range of the parameter space. Finding the
%orthonormal dual space allows the inner product to be redefined
%using the complexification procedure of Bender for non-Hermitian
%Hamiltonians. Transition probabilities governed by H are shown to be
%manifestly unitary when the complexification procedure is followed.
%A specific transition element between harmonic oscillator states is
%evaluated for both the Hermitian and non-Hermitian cases to identify
%the differences in time evolution. (C) 2004 American Institute of
%Physics

A. Mostafazadeh, J. Phys. A: Math. Gen. 39 (2006) 10171;

%in the coordinate basis {|xi}, we find
A. Mostafazadeh, J. Math. Phys. 47 (2006) 072103;

%delta
A. Mostafazadeh, J. Phys. A: Math. Gen. 39 (2006) 13506;

%Moyal :
F. G. Scholtz and H. B. Geyer, Phys. Lett. B 634 (2006) 84;
%
%
%Operator equations and Moyal products-metrics in quasi-Hermitian quantum mechanics
%By: Scholtz, FG; Geyer, HB
%PHYSICS LETTERS B  Volume: 634   Issue: 1   Pages: 84-92   Published: MAR 2 2006
% Full Text from Publisher Full Text View AbstractClose Abstract Times Cited: 25
%(from All Databases
%
%Operator equations and Moyal products-metrics in quasi-Hermitian quantum mechanics
%By: Scholtz, FG; Geyer, HB
%PHYSICS LETTERS B  Volume: 634   Issue: 1   Pages: 84-92   Published: MAR 2 2006
% Full Text from Publisher Full Text View AbstractClose
%
%
% Abstract
%The Moyal product is used to cast the equation for the metric of a
%non-Hermitian Hamiltonian in the form of a differential equation.
%For Hamiltonians of the form p(2) + V(ix) with V polynomial this is
%an exact equation. Solving this equation in perturbation theory
%recovers known results. Explicit criteria for the hermiticity and
%positive definiteness of the metric are formulated on the functional
%level. (c) 2006 Elsevier B.V. All rights reserved.


%Moyal
C. Figueira de Morisson Faria and A. Fring, Czech J. Phys. 56 (2006)
899;


%Pert m,ethod:
C. Figueira de Morisson Faria and A. Fring, J. Phys. A: Math. Gen.
39 (2006) 9269;

%5. Choice of a metric for the non-Hermitian oscillator By:
P. Musumbu-Dibwe, H. G. Geyer, W. D. Heiss,
 J. Phys. A: Math. Gen. 40 (2007) F75;
%D. P.; Geyer, H. B.; Heiss, W. D. JOURNAL OF PHYSICS A-MATHEMATICAL
%AND THEORETICAL  Volume: 40 Issue: 2   Pages: F75-F80   Published:
%JAN 12 2007
% Full Text from Publisher Full Text View AbstractClose Abstract
%
%The harmonic oscillator Hamiltonian, when augmented by a
%non-Hermitian PT-symmetric part, can be transformed into a Hermitian
%Hamiltonian. This is achieved by introducing a metric which, in
%general, renders other observables such as the usual momentum or
%position as non-Hermitian operators. The metric depends on one real
%parameter, the full range of which is investigated. The explicit
%functional dependence of the metric and each associated Hamiltonian
%is given. A specific choice of this parameter determines a specific
%combination of position and momentum as being an observable; this
%can be in particular either standard position or momentum, but not
%both simultaneously. Singularities of the metric are explored and
%their removability is investigated. The physical significance of
%these findings is discussed.

C. Quesne, J. Phys. A: Math. Gen. 40 (2007) F745;
%Lie:
%
%
%A non-Hermitian oscillator Hamiltonian and su(1,1): a way towards
%generalizations By: Quesne, C. JOURNAL OF PHYSICS A-MATHEMATICAL AND
%THEORETICAL  Volume: 40   Issue: 30   Pages: F745-F751   Published:
%JUL 27 2007
% Full Text from Publisher Full Text View AbstractClose
%
%  Abstract
%The family of metric operators, constructed by Musumbu et al (2007
%J. Phys. A: Math. Theor. 40 F75), for a harmonic oscillator
%Hamiltonian augmented by a non- Hermitian PT-symmetric part, is
%re-examined in the light of an su(1,1) approach. An alternative
%derivation, only relying on properties of su(1,1) generators, is
%proposed. Being independent of the realization considered for the
%latter, it opens the way towards the construction of generalized
%non- Hermitian (not necessarily PT-symmetric) oscillator
%Hamiltonians related by similarity to Hermitian ones. Some examples
%of them are reviewed.Times Cited: 17 (from All Databases)

% 50. Non-Hermitian Hamiltonians with real eigenvalues coupled to electric fields:
% From the time-independent to the time-dependent quantum mechanical formulation
%By: Faria, C. F. M.; Fring, A. LASER PHYSICS  Volume: 17   Issue: 4
%Pages: 424-437   Published: APR 2007
% Full Text from Publisher Full Text View AbstractClose
%
%  Abstract
%We provide a reviewlike introduction to the quantum mechanical
%formalism related to non-Hermitian Hamiltonian systems with real
%eigenvalues. Starting with the time-independent framework, we
%explain how to determine an appropriate domain of a non-Hermitian
%Hamiltonian and pay particular attention to the role played by PJ
%symmetry and pseudo-Hermiticity. We discuss the time evolution of
%such systems having in particular the question in mind of how to
%couple consistently an electric field to pseudo-Hermitian
%Hamiltonians. We illustrate the general formalism with three
%explicit examples: (i) the generalized Swanson Hamiltonians, which
%constitute non-Hermitian extensions of anharmonic oscillators, (ii)
%the spiked harmonic oscillator, which exhibits explicit
%super-symmetry, and (iii) the -x(4)-potential, which serves as a toy
%model for the quantum field theoretical phi(4)-theory.
%
P. E. G. Assis and A. Fring, J. Phys. A: Math. Theor. 41 (2008)
244001;



C. Quesne, J. Phys. A: Math. Theor. 41 (2008)  244022;
%
%Quasi-Hermitian supersymmetric extensions of a non-Hermitian
%oscillator Hamiltonian and of its generalizations By: Quesne, C.
%Conference: 6th International Workshop on Pseudo-Hermitian
%Hamiltonians in Quantum Physics Location: City Univ London, London,
%ENGLAND Date: JUL 16-18, 2007 Sponsor(s): London Math Soc; Inst
%Phys; Doppler Inst; City Univ London, Sch Engn & Math Sci JOURNAL OF
%PHYSICS A-MATHEMATICAL AND THEORETICAL  Volume: 41   Issue: 24
%Article Number: 244022   Published: JUN 20 2008
% Full Text from Publisher Full Text View AbstractClose
%
%  Abstract
%A harmonic oscillator Hamiltonian augmented by a non-Hermitian
%PT-symmetric part and its su(1,1) generalizations, for which a
%family of positive-definite metric operators was recently
%constructed, are re-examined in a supersymmetric context. Some
%quasi-Hermitian supersymmetric extensions of such Hamiltonians are
%proposed by enlarging su(1,1) to a su(1, 1/1) similar to osp(2/ 2,
%R) superalgebra. This allows the construction of new non-Hermitian
%Hamiltonians related by similarity to Hermitian ones. Some examples
%of them are reviewed.





P. E. G. Assis and A. Fring, J. Phys. A: Math. Theor. 42 (2009)
015203;
%
%
%Non-Hermitian Hamiltonians of Lie algebraic type
%By: Assis, Paulo E. G.; Fring, Andreas
%JOURNAL OF PHYSICS A-MATHEMATICAL AND THEORETICAL  Volume: 42   Issue: 1
% Article Number: 015203   Published: JAN 9 2009
% Full Text from Publisher Full Text View AbstractClose
%  Abstract
%We analyse a class of non-Hermitian Hamiltonians, which can be
%expressed in terms of bilinear combinations of generators in the
%sl(2)(R)-Lie algebra or their isomorphic su(1, 1)-counterparts. The
%Hamiltonians are prototypes for solvable models of Lie algebraic
%type. Demanding a real spectrum and the existence of a well-defined
%metric, we systematically investigate the constraints these
%requirements impose on the coupling constants of the model and the
%parameters in the metric operator. We compute isospectral Hermitian
%counterparts for some of the original non-Hermitian Hamiltonians.
%Alternatively, we employ a generalized Bogoliubov transformation,
%which allows us to compute explicitly real-energy eigenvalue spectra
%for these type of Hamiltonians, together with their eigenstates. We
%compare the two approaches
%
%
%Metrics and isospectral partners for the most generic
%cubic PT-symmetric non-Hermitian Hamiltonian
%By: Assis, Paulo E. G.; Fring, Andreas
%Conference: 6th International Workshop on Pseudo-Hermitian Hamiltonians
%in Quantum Physics Location: City Univ London, London, ENGLAND Date: JUL 16-18, 2007
%Sponsor(s): London Math Soc; Inst Phys; Doppler Inst; City Univ London,
%Sch Engn & Math Sci
%JOURNAL OF PHYSICS A-MATHEMATICAL AND THEORETICAL  Volume: 41   Issue: 24
%Article Number: 244001   Published: JUN 20 2008
% Full Text from Publisher Full Text View AbstractClose
%  Abstract
%We investigate properties of the most general PT-symmetric
%non-Hermitian Hamiltonian of cubic order in the annihilation and
%creation operators as a ten-parameter family. For various choices of
%the parameters we systematically construct an exact expression for a
%metric operator and an isospectral Hermitian counterpart in the same
%similarity class by exploiting the isomorphism between operator and
%Moyal products. We elaborate on the subtleties of this approach. For
%special choices of the ten parameters the Hamiltonian reduces to
%various models previously studied, such as to the complex cubic
%potential, the so-called Swanson Hamiltonian or the transformed
%version of the from below unbounded quartic -x(4)-potential. In
%addition, it also reduces to various models not considered in the
%present context, namely the single-site lattice Reggeon model and a
%transformed version of the massive sextic +/- x(6)-potential, which
%plays an important role as a toy model to identify theories with
%vanishing cosmological constant.
%


H. F. Jones, J. Phys. A: Math. Theor. 42 (2009) 135303;
%
%Gauging non-Hermitian Hamiltonians
%By: Jones, H. F.
%JOURNAL OF PHYSICS A-MATHEMATICAL AND THEORETICAL  Volume: 42   Issue: 13
%Article Number: 135303   Published: APR 3 2009
% Full Text from Publisher Full Text View AbstractClose
%  Abstract
%We address the problem of coupling non-Hermitian systems, treated as
%fundamental rather than effective theories, to the electromagnetic
%field. In such theories the observables are not the x and p
%appearing in the Hamiltonian, but quantities X and P constructed by
%means of the metric operator. Following the analogous procedure of
%gauging a global symmetry in Hermitian quantum mechanics we find
%that the corresponding gauge transformation in X implies minimal
%substitution in the form P -> P - eA(X). We discuss how the relevant
%matrix elements governing electromagnetic transitions may be
%calculated in the special case of the Swanson Hamiltonian, where the
%equivalent Hermitian Hamiltonian  h is local, and in the more
%generic example of the imaginary cubic interaction, where H is local
%but h is not.



S. Dey, A. Fring and B, Khantoul, J. Phys. A: Math. Theor. 46 (2013)
335304;
%
%
%Hermitian versus non-Hermitian representations for minimal length
%uncertainty relations By: Dey, Sanjib; Fring, Andreas; Khantoul,
%Boubakeur JOURNAL OF PHYSICS A-MATHEMATICAL AND THEORETICAL  Volume:
%46   Issue: 33     Article Number: 335304   Published: AUG 23 2013
% Full Text from Publisher Full Text View AbstractClose
%
% Abstract
%We investigate four different types of representations of deformed
%canonical variables leading to generalized versions of Heisenberg's
%uncertainty relations resulting from noncommutative spacetime
%structures. We demonstrate explicitly how the representations are
%related to each other and study three characteristically different
%solvable models on these spaces, the harmonic oscillator, the
%manifestly non-Hermitian Swanson model and an intrinsically
%noncommutative model with Poschl-Teller type potential. We provide
%an analytical expression for the metric in terms of quantities
%specific to the generic solution procedure and show that when it is
%appropriately implemented expectation values are independent of the
%particular representation. A recently proposed inequivalent
%representation resulting from Jordan twists is shown to lead to
%unphysical models. We suggest an anti-PT-symmetric modification to
%overcome this shortcoming.
%
%%
%
% arXiv:0910.5017 [pdf]
%Title: PT-Invariance and Indefinite Metric Authors: Scott Chapman
%
%A new proof is given for why the non-Hermitian, PT-Invariant cubic
%oscillator with imaginary coupling has real eigenvalues. The proof
%consists of two steps. In the first step, it is shown that for many
%PT-Invariant Hamiltonians, one can define corresponding Hermitian
%Hamiltonians that have the same eigenvalues when quantized using
%indefinite metric. The second step is to show that the
%indefinite-metric eigenvalues of the Hermitian cubic oscillator are
%real since the norms of its eigenstates do not vanish. The
%correspondence between PT-Invariant Hamiltonians and
%indefinite-metric Hermitian Hamiltonians is further discussed as a
%way to determine vacuum stability for certain supersymmetric gauge
%theories.


A. Ghatak and B. P. Mandal,
%Comparison of different approaches of finding the positive
%definite metric in pseudo-Hermitian theories,
Comm. Theor. Phys. 59 (2013) 533.

\bibitem{BG}
V. Buslaev and V. Grechi,
%Equivalence of unstable anharmonic
%oscillators and double wells,
J. Phys. A: Math. Gen. 26 (1993) 5541.
%-5549, 1993.

\bibitem{chain}
%\cite{Korff}
%\bibitem{Korff}
% arXiv:math-ph/0703085 [pdf, ps, other]
%Title: PT Symmetry on the Lattice: The Quantum Group Invariant XXZ
%Spin-Chain Authors: Christian
Ch. Korff and  R. A. Weston,
% Comments: 32
%pages with figures, v2: some minor changes and added references,
%version published in JPA Journal-ref:
J. Phys. A: Math. Theor. 40
%-8872,
(2007) 8845;
%

M. Znojil, %Maximal couplings in PT-symmetric chain-models with the
%real spectrum of energies.
J. Phys. A: Math. Theor. 40  (2007) 4863;
%4875. (mathph/ 0703070).


O. A. Castro-Alvaredo and A. Fring,
%A spin chain model with
%non-Hermitian interaction: the Ising quantum spin chain in an
%imaginary field,
J. Phys. A: Math. Theor. 42 (2009) 465211;

O. Bendix, R. Fleischmann, T. Kottos and B. Shapiro, Phys. Rev.
Lett. 103 (2009) 030402;


L. Jin and Z. Song,
%Wave-packet dynamics in a non-Hermitian PT symmetric tight-binding
%chain,
Comm. Theor. Phys. 54 (2010) 73;

M. Znojil,
%Gegenbauer-solvable quantum chain model,
Phys. Rev. A 82 (2010) 052113;

S. Longhi,  Phys. Rev. A 88 (2013) 052102;

%PT-Symmetry in Non-Hermitian Su-Schrieffer-Heeger
%model with complex boundary potentials
%May 23, '14 3:21 PM 3 June
B.-G. Zhu, R. Lu and S. Chen, Phys. Rev. A  89 (2014) 062102.


\bibitem{feasiblesdis}
%M. Znojil,
%Complete set of inner products for a discrete PT-symmetric square-well Hamiltonian,
%J. Math. Phys. 50, 122105, 2009.
%
M. Znojil,
%Complete set of inner products for a discrete
%PT-symmetric square-well Hamiltonian.
J. Math. Phys. 50 (2009) 122105;
%, http://dx.doi.org/10.1063/1.3272002, (arXiv:0911.0336v1
%[math-ph] 2 Nov 2009)

E. Ergun and M. Saglam, Rep. Math. Phys. 65 (2010) 367.
%Volume 65, Issue 3, June 2010, Pages 367–378
% On the metric of a non-Hermitian model
% Department of Physics, Ankara University, 06100 Tandogan, Ankara, Turkey
%Received 19 October 2009, Revised 2 November 2009, Available online 7 October 2010
%  Show more
%
%%
%Reports on Mathematical Physics
%Volume 65, Issue 3, June 2010, Pages 367–378
% On the metric of a non-Hermitian model
%Ebru Ergun , Mesude Saglam Department of Physics,
%Ankara University, 06100 Tandogan, Ankara, Turkey
%Received 19 October 2009, Revised 2 November 2009, Available online 7 October 2010
%  Show more


\bibitem{feasiblesdisbe}
M. Znojil,
%Cryptohermitian picture of scattering using quasilocal
%metric operators SYMMETRY, INTEGRABILITY and GEOMETRY: METHODS and
%APPLICATIONS
SIGMA 5 (2009) 085.
% 21 pages , (doi:
%doi:10.3842/SIGMA.2009.085); (arXiv:0908.4045 [quant-ph math-ph
%math.MP] 27 Aug 2009) a part of the special issue associated with
%the DI microconference ''Analytic and algebraic methods V'' (May 27
%- 28, 2009, 9.30 a.m. - 5.00 p.m., Villa Lanna, Prague)

\bibitem{Bila}
H. B\'{\i}la,
%H.: Non-Hermitian Operators in Quantum Physics.
PhD thesis, Charles University (2008) (listing older references);

H. B\'{\i}la,
 %17. Bíla, H.:
 Adiabatic
time-dependent metrics in PT-symmetric quantum theories.
arXiv:0902.0474 (unpublished).

\bibitem{Fluegge}
A. Messiah, Quantum Mechanics, North Holland, Amsterdam, 1961.


\bibitem{disqw}
M. Znojil,
%Discrete quantum square well of the first kind"
Phys. Lett.
A 375 (2011) 2503.
%
% - 2509 http://dx.doi.org/10.1016/j.physleta.2011.05.027
%arXiv:1105.1863


\bibitem{chainb}
L. Jin and Z. Song, Phys. Rev. A 80 (2009) 052107.

\bibitem{chaince}
Y. N. Joglekar, D. Scott, M. Babbey and A. Saxena, Phys. Rev. A 82
(2010) 030103(R).

\bibitem{Langer}
H. Langer and Ch. Tretter,
%A Krein space approach to PT symmetry,
Czech. J. Phys. 54 (2004) 1113;
%-1120,.

P. Siegl,
%PT-Symmetric Square Well-Perturbations and the Existence of Metric Operator.
Int. J. Theor. Phys.  50 (2011) 991.
%–996, 2011.


\bibitem{sqw}
M. Znojil,
%PT symmetric square well (quant-ph/0101131),
Phys. Lett. A 285 (2001) 7;
%- 10.

M. Znojil and G. L\'{e}vai,
%Spontaneous breakdown of PT symmetry in the solvable square well model
%(hep-th/0111213)
Mod. Phys. Letters A 16 (2001) 2273;
% - 2280.

M. Znojil, %Coupled-channel version of PT-symmetric square well,
J. Phys. A: Math.
Gen. 39 (2006) 441.
%-455, 2006

\bibitem{Batal}
A. Mostafazadeh and A. Batal, J. Phys. A: Math. Theor. 37  (2004)
11645.
%
% arXiv:quant-ph/0408132 [pdf, ps, other]
%Title: Physical Aspects of Pseudo-Hermitian and $PT$-Symmetric
%Quantum Mechanics Authors: Ali Mostafazadeh, Ahmet Batal Comments:
%45 pages, 13 figures, 2 tables Journal-ref: J.Phys. A37 (2004)
%11645-11680
%

\bibitem{sdavidem}
D. Krejcirik, H. Bila and M. Znojil,
%"Closed formula for the metric in the Hilbert space of a PT-symmetric model";
J. Phys. A 39 (2006) 10143.
%-10153.
%Available on [ IoP (published version) ].
%Preprint available on [ math-ph/0604055 ].
%Citations: 26

\bibitem{Zelezny}
J. Zelezny, Int. J. Theor. Phys. 50 (2011) 1012;
%Zelezn´y, J. The Krein-space theory for non-Hermitian PT -symmetric operators.
%Master's thesis, FNSPE, CTU in Prague, 2010/2011.
%[35] ¡Zelezn´y, J. Spectrum of the Metric Operator of a Simple PT -Symmetric Model.
%International Journal of Theoretical Physics 50 (2011), 1012–1018.

%  On the similarity of Sturm-Liouville operators
%  with non-Hermitian boundary conditions to self-adjoint and normal operators
D. Krejcirik, P. Siegl, and J. Zelezny,
Compl. Anal. Oper. Theory 8 (2014) 255.
%-281 arXiv:1108.4946

%D. Krejcirik, P. Siegl and J. Zelezny:
%"On the similarity of Sturm-Liouville operators with non-Hermitian
%boundary conditions to self-adjoint and normal operators";
%Complex Anal. Oper. Theory 8 (2014), 255-281.
%Available on [ Springer (published version) ].
%Preprint on arXiv:1108.4946 [math.SP].




\bibitem{fragile}
M. Znojil,
%Fragile PT-symmetry in a solvable model.
J. Math. Phys. 45 (2004) 4418.
% - 30 (math-ph/0403033),



\bibitem{Tater}
G. L\'{e}vai and M. Znojil,
%Conditions for complex spectra in a class of PT symmetric potentials
%(quant-ph/0110064)
Mod. Phys. Letters A 16 (2001) 1973;
% - 1981

 % Pseudospectra in non-Hermitian quantum mechanics
D. Krejcirik, P. Siegl, M. Tater and J. Viola, submitted
(arXiv:1402.1082).

\bibitem{Ryshik}
M. Abramowitz and I. A. Stegun, Handbook of Mathematical Functions,
Dover, New York, 1970.

%NIST Digital Library of Mathematical Functions, Chapter 18, Orthogonal Polynomials:
%http://dlmf.nist.gov/18

\bibitem{feasibles}
D. Krejcirik,
%"Calculation of the metric in the Hilbert space
%of a PT-symmetric model via the spectral theorem".
J. Phys. A: Math. Theor. 41 (2008) 244012.
%Available on [ IoP (published version) ].
%Preprint available on arXiv:0707.1781 [math-ph].
%Citations: 19
%
%
%Calculation of the metric in the Hilbert space of a PT-symmetric
%model via the spectral theorem By: Krejcirik, David Conference: 6th
%International Workshop on Pseudo-Hermitian Hamiltonians in Quantum
%Physics Location: City Univ London, London, ENGLAND Date: JUL 16-18,
%2007 Sponsor(s): London Math Soc; Inst Phys; Doppler Inst; City Univ
%London, Sch Engn & Math Sci JOURNAL OF PHYSICS A-MATHEMATICAL AND
%THEORETICAL  Volume: 41   Issue: 24     Article Number: 244012
%Published: JUN 20 2008
%
%In a previous paper [1] we introduced a very simple PT-symmetric
%non-Hermitian Hamiltonian with a real spectrum and derived a closed
%formula for the metric operator relating the problem to a Hermitian
%one. In this paper we propose an alternative formula for the metric
%operator, which we believe is more elegant and whose
%construction-based on a backward use of the spectral theorem for
%self-adjoint operators-provides new insights into the nature of the
%model


\bibitem{sample}
M. Znojil,
%Znojil, M.:
J. Phys. Conf. Ser. 343 (2012) 012136;

M. Znojil,
%Miloslav Znojil, "N-site-lattice analogues of V(x)=i x^3"
Ann. Phys. (NY) 327 (2012) 893.
% - 913
%http://dx.doi.org/10.1016/j.aop.2011.12.009 arXiv: 1111.0484




\end{thebibliography}
\end{document}